# Coherent all-optical tuning of large-area phase-gradient metasurface

ZHIPING HE,[1,3] XU FANG,[2,*] AND JUEJUN HU[3]

[1]*Department of Electrical Engineering and Computer Science, Massachusetts Institute of Technology, Cambridge, Massachusetts 02139, USA*
[2]*School of Electronics and Computer Science, University of Southampton, SO17 1BJ, UK*
[3]*Department of Materials Science and Engineering, Massachusetts Institute of Technology, Cambridge, Massachusetts 02139, USA*
*[*]x.fang@soton.ac.uk*

**Abstract:** Tunable active metasurfaces have become a major research focus in recent years. Among tuning mechanisms, all-optical coherent control stands out because it requires no material or geometric change, enabling ultrafast, low-energy, interference-based modulation of amplitude, phase, and polarization in ultrathin devices. However, when applied to phase-gradient metasurfaces, coherent control has been limited to small apertures effectively confined to a single Fresnel zone, leading to large divergence and degraded beam quality. Here we propose and numerically validate a scalable method that enables large-area coherent control. The key idea is to use coherent illumination to tune the phase gradient within each Fresnel zone while a direct search algorithm optimizes zone-by-zone parameters to meet system-level targets. Using this principle, we demonstrate continuous tuning of a large-area metasurface for continuous beam-steering without per-meta-atom phase actuation. The same framework applies broadly to continuously tunable phase-gradient optics, including varifocal metalenses, parfocal zoom metalenses, tunable axicons, and related dynamic focusing elements.

## 1. Introduction

Active metasurfaces can be reconfigured through a variety of mechanisms, including thermo-optic heating [1,2], electro-optic effects [3,4], free-carrier injection [5,6], liquid-crystal reorientation [7–9], phase transition in various materials [10–13], and mechanical actuation via MEMS [14] or elastomeric strain [15]. Among these, coherent control uses interference between phase-locked beams to program an otherwise static nanostructure, tuning effective amplitude, phase, polarization, and absorption without altering geometry or material properties [16–20]. It offers ultrafast response (set by optical phase rather than charge transport) and low energy overhead since no Joule heating or physical motion is required.

However, when coherent control is applied to phase-gradient metasurfaces, prior demonstrations have been restricted to apertures effectively within a single Fresnel zone [21–23], which leads to strong beam divergence and degraded quality along the gradient direction. In principle, one might attempt to extend the design periodically to cover a larger area, treating a small coherent-control unit as a repeating supercell. Yet this periodic extension inevitably enforces discrete translational symmetry, which confines the optical response to a set of quantized diffraction orders defined by the grating equation. As a result, the metasurface loses the defining advantage of coherent control—its ability to tune the output angle continuously by varying the phase relation between counter-propagating beams. Instead of smooth beam steering, the output becomes locked to discrete angles associated with each diffraction order, preventing continuous modulation of the far-field direction and fundamentally limiting scalability.

To address these limitations, we introduce a scalable coherent-control framework based on the concept of a *supercell*—a repeating coherent-control cell that forms the building block of the metasurface. While the metasurface still consists of periodically arranged supercells, we employ spatial light modulators (SLMs) to modulate both the phase and amplitude of the light incident on each supercell. This configuration enables independent tuning of the phase gradient within every supercell, effectively breaking the global periodic symmetry that otherwise constrains the optical response to discrete diffraction orders. Through direct search optimization of the gradient in each supercell [24], we achieve continuous tuning across a large-aperture metasurface while maintaining exceptional beamforming quality.

Using beam steering as an example, we demonstrate that this approach yields a substantially wider angular tuning range and higher diffraction efficiency than direct SLM-based modulation of phase and amplitude across each supercell without the coherent-control mechanism. The significance of this result is two-fold. First, it provides a practical route to overcoming the aperture-size limitation inherent in traditional coherent-control metasurfaces, thereby extending coherent control from uniform modulation to spatially varying, phase-gradient modulation—a far more powerful mode of optical functionality. Second, it defies the conventional wisdom that individual control of every meta-atom is essential for continuous tuning [25]. In a large-aperture phase-gradient metasurface, the boundaries between adjacent Fresnel zones shift continuously as the output angle varies. Each time a boundary crosses a meta-atom, a discrete $2\pi$ phase jump must be maintained to preserve constructive interference in the far field. This condition dictates that every meta-atom be capable of independent phase adjustment to follow the moving zone structure. Extending the achievable phase coverage beyond $2\pi$ (as practiced in dispersion-engineered metasurfaces [26,27]) can partially alleviate this constraint, but the practical range of phase delay remains limited, imposing the same small-aperture restriction observed in achromatic metasurface implementations [28]. By contrast, our approach tunes the phase gradient within each supercell rather than the absolute phase of individual meta-atoms, enabling continuous large-area control through far simpler and experimentally accessible parameters, such as those available from SLM-based modulation. This phase-gradient modulation strategy is broadly compatible with other active tuning mechanisms [29,30] and can similarly overcome aperture-related constraints to realize continuously tunable phase-gradient metasurface optics.

## 2. Method

### 2.1 Device architecture

We adopted the meta-atom design reported by He *et al.* to leverage the phase gradient generated by the dispersion contrast between the electric and magnetic dipole resonances of silicon nanopillars [22]. Figure 1a illustrates a single supercell, which serves as the fundamental repeating unit in the large-area metasurface. Each supercell consists of a 10 × 10 array of silicon nanopillars fabricated on a glass substrate, each 160 nm in height with a lattice period of 300 nm along both *x* and *y* axes. The nanopillars are identical along the *y* direction, while their diameters (*D*) increase from 120 nm to 174 nm in 6 nm increments along *x*, thereby producing a controllable phase gradient in that direction. The resulting supercell, with lateral dimensions of 3 μm × 3 μm, is periodically replicated to form the full metasurface, enabling a large optical aperture.

The overall configuration of the beam steering system is shown in Fig. 1b, which is adapted from the work by He *et al.* with the only addition of SLMs. A linearly *x*-polarized input beam with a free-space wavelength of $\lambda_0 = 555$ nm, is first directed onto a phase-only spatial light modulator (SLM 1) with a pixel pitch of 3 μm × 3 μm, corresponding one-to-one with the metasurface supercells. The modulated beam then passes through a beam splitter that divides the incident light into two beams, which are subsequently directed by mirrors to propagate in opposite directions. The upward-propagating component (+*z*) is further modulated by a bipolar-amplitude spatial light modulator (SLM 2), which has the same pixel pitch as SLM 1 and is also aligned with the supercell lattice, before recombining with the downward-propagating component (−*z*) at the metasurface. The interference of these counter-propagating beams produces metasurface response that depends on the coherent interference condition between the two beams, thereby enabling beam steering into the designed output direction. In this configuration, only two parameters are adjusted per supercell: the relative phase among supercells $\varphi_0$ and the magnetic field ratio between the substrate and free-space incident beams at the *x-y* plane that bisecting the nanopillars at mid height ($B_s/B_f$), the latter of which is controlled by relative intensities of the two beams. This reduced parameter space significantly improves the feasibility and scalability of the proposed method, as we will show later.

We simulated the optical response of individual meta-atoms using the Lumerical finite-difference time-domain (FDTD) solver. Following He *et al.*, the refractive index of the amorphous silicon nanopillars was taken to be 4.06 + 0.03i, and that of the glass substrate to be 1.5. Periodic boundary conditions were along the *x* and *y* axes. The simulated responses are presented in Fig. 1c, which are in close agreement with those reported by He *et al.*, where COMSOL Multiphysics was employed. This confirms that a supercell incorporating a diameter gradient of silicon nanopillars yields a continuously tunable phase gradient via coherent control. This property is essential for achieving continuous beam-angle steering.

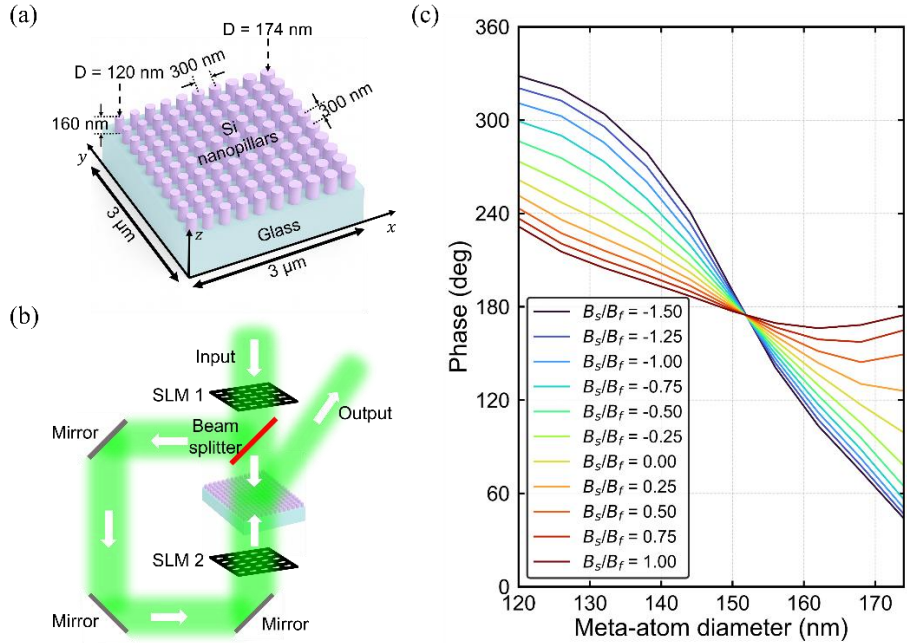

Fig. 1. (a) Schematic illustration of the supercell of the tunable phase-gradient metasurface realized through coherent control. The meta-atoms, composed of silicon nanopillars with uniform height, are identical along the *y* direction and exhibit a diameter gradient along the *x* direction. (b) Conceptual schematic of the proposed beam-steering system. SLM 1 denotes a phase-only spatial light modulator, and SLM 2 denotes a bipolar amplitude spatial light modulator. (c) Calculated dependence of the phase shift of the output beam, relative to the free-space incident beam, on the meta-atom diameter and the magnetic field ratio ($B_s/B_f$).

## 2.2 Direct search optimization algorithm

We employed the direct search algorithm [24,31] to identify the design that maximizes the transmitted energy in the desired beam-steering direction while suppressing undesired diffraction lobes. In particular, we optimize the light intensity calculated from the Fresnel-Kirchhoff diffraction integral under the Fraunhofer approximation, expressed as:

$$U(\theta_x, \theta_y) \propto \iint t(x_0, y_0) \exp\left(-ik(x_0 \sin\theta_x + y_0 \sin\theta_y)\right) dx_0 dy_0,$$

$$I(\theta_x, \theta_y) = |U(\theta_x, \theta_y)|^2,$$

where $U(\theta_x, \theta_y)$ denotes the diffracted field in the direction $(\theta_x, \theta_y)$ and $I(\theta_x, \theta_y)$ represents the corresponding light intensity. $(\theta_x, \theta_y)$ are the diffraction angles along the *x* and *y* directions, such that the unit vector towards the output beam direction is

$\left(\sin\theta_x, \sin\theta_y, \sqrt{1-\sin^2\theta_x-\sin^2\theta_y}\right)$. Here $k = 2\pi/\lambda_0$ is the free-space wavenumber, and $(x_0, y_0)$ are the spatial coordinates on the metasurface. The term $t(x_0, y_0)$ is the complex transmission coefficient that encodes both the phase modulation and amplitude response of the pixel located at $(x_0, y_0)$. In this formulation, we account for both the transmission through the metasurface and the amplitude modulation introduced by SLM 2.

The figure of merit (FOM) for the direct search optimization is defined to suppress unwanted diffraction orders and thereby enhance beam-steering efficiency. Specifically, the FOM penalizes power leakage into the spurious diffraction orders along the $x$ direction. The FOM is expressed as:

$$\text{FOM} = I(\theta_x, \theta_y) - w \times \left(I(\theta_{x,-1}, \theta_y) + I(\theta_{x,+1}, \theta_y)\right).$$

$(\theta_x, \theta_y)$ is the desired beam steering direction, while $\theta_{x,-1}$ and $\theta_{x,+1}$ correspond to the $-1$ and $+1$ diffraction orders (as gauged based on the supercell period), respectively. The weighting parameter $w$ determines the relative strength of the applied penalty, and in the simulations presented here we empirically set $w = 5.0$. Unless otherwise specified, we restrict our analysis to $\theta_y = 0$, corresponding to beam steering exclusively along the $x$ direction, with the exception of Section 3.4, where two-dimensional beam steering in both $\theta_x$ and $\theta_y$ is investigated. We also assume $(\varphi_0, B_s/B_f)$ is invariant along the $y$ axis when we simulate the $\theta_y = 0$ case.

The direct search algorithm was subsequently used to optimize two design parameters in each supercell: $(\varphi_0, B_s/B_f)$. The relative phase was selected from the interval $[0, 2\pi]$ with 51 uniformly spaced discrete levels. Likewise, $B_s/B_f$ was sampled from the range $[-1.5, 1.0]$ (following He *et al.*) with 51 levels. As a result, a library of $51 \times 51$ candidate pairs $(\varphi_0, B_s/B_f)$ was constructed for each supercell. The optimization began by randomly assigning an initial pair to each supercell. The algorithm then iteratively updated the configuration: for each unoptimized supercell, all $51 \times 51$ candidate pairs were exhaustively tested, and the pair yielding the maximum FOM was retained. This procedure was repeated for all supercells in a random sequence during each iteration and terminated when the relative improvement in the FOM after one iteration was less than 0.01%.

### 3. Results and discussion

#### 3.1 Continuous beam steering performance

In this section, we analyze the performance of a square metasurface with an aperture size of 150 μm × 150 μm, designed according to the method described above. The characterization focuses on the relative diffraction efficiency, here defined as the ratio of the optical power directed into the desired primary diffraction order to the total transmitted power into free space [32]. It serves as a critical system-level parameter because it directly reflects the signal-to-noise ratio. The optical power in the target order is obtained by integrating the intensity distribution over a square region centered at the maximum intensity and with an edge length equal to five times the angular full-width-at-half-maximum (FWHM) of the main lobe. Figure 2a shows the relative diffraction efficiency for the $-1$ (blue squares), primary (green circles), and $+1$ (red diamonds) diffraction orders. The results indicate that the relative diffraction efficiency of the primary order remains above 60% across a steering range from approximately 1° to 12°. This observation confirms that large-aperture beam steering is achieved over roughly the same angular range as reported for micron-scale aperture metasurface by He *et al.* [22]. As illustrated in Fig. 2a, the relative diffraction efficiency of the primary diffraction order decreases rapidly at larger angles. This is because the range of accessible phase gradients is limited, and according to the generalized Snell's law, the maximum allowed steering angle is $\theta_x = \arcsin(\lambda_0/2\pi \times d\varphi/dx)$, where $d\varphi/dx$ denotes the maximum phase gradient attainable in each supercell. This accounts for the decay in primary diffraction order efficiency when $\theta_x$ exceeds approximately 12°.

Figures 2b-e show the intensity profiles as a function of $\theta_x$ for a set of representative steering angles, along with a comparison to the results reported by He *et al.*. Two key observations can be drawn. First, the beam width of the primary diffraction order is substantially smaller than that reported by He *et al.*, demonstrating that our proposed design significantly reduces beam divergence and improves beam quality relative to the earlier approach. For the diffraction angles of 1.5°, 5.7°, 8.1°, and 10.6°, the angular FWHM of the primary diffraction order is 0.194°, 0.196°, 0.196°, and 0.196°, respectively, which matches closely with the diffraction-limited beam divergence of 0.188°, 0.189°, 0.190°, and 0.191°, respectively. Second, the peak intensity of the unwanted diffraction orders is much lower than that of the desired lobe, confirming that the metasurface design effectively suppresses aliasing and concentrates the majority of the transmitted power into the intended direction. This outcome is consistent with the high relative diffraction efficiency observed in Fig. 2a. Our results therefore confirm that the new design can offer efficient, continuous beam steering with diffraction-limited beam quality.

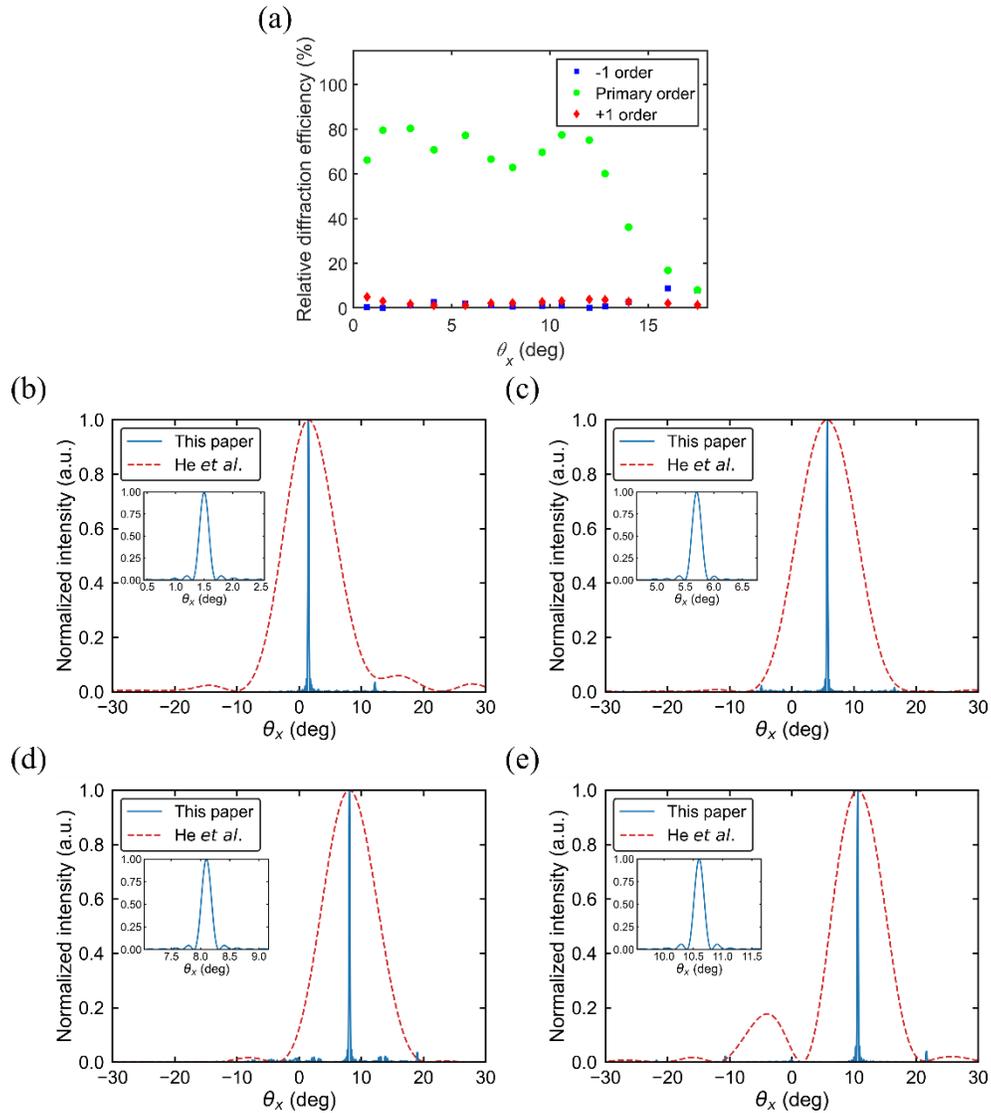

Fig. 2. (a) Relative diffraction efficiencies of the −1 (blue squares), primary (green circles), and +1 (red diamonds) diffraction orders of the phase-gradient metasurface. (b-e) Normalized intensity profiles as a function of $\theta_x$ for steering angles of (b) 1.5°, (c) 5.7°, (d) 8.1°, and (e) 10.6°. The solid blue curves represent our results, and the red dashed curves correspond to the results reported by He *et al.* [22]. In both cases, data are normalized to the respective maximum intensity. Insets in (b-e) provide magnified views of the primary diffraction order for our proposed design.

## 3.2 Scaling to large aperture size

In this section, we demonstrate that the high performance achieved for the 150 μm-wide metasurface, characterized by diffraction-limited beam quality and high relative diffraction efficiency, can be seamlessly scaled to larger apertures while maintaining these key attributes. Figure 3 illustrates the variation of the beam steering performance with the metasurface size (aperture size). The blue squares, green circles, and red diamonds correspond to the data for the beam steered towards angles of 1.5°, 5.7°, and 9.6°, respectively, while the black curve denotes the theoretical diffraction limit. As shown in Fig. 3a, the angular FWHM along the $\theta_x$ direction of the primary diffraction lobe consistently approaches the diffraction limit across the examined aperture sizes, demonstrating that the algorithm can be generalized to larger apertures while maintaining beam divergence close to the diffraction-limited condition. Figure 3b depicts the relative diffraction efficiency of the primary diffraction order as a function of the aperture size. Over the scanned range from 30 μm to 240 μm, the relative diffraction efficiency remains essentially constant, thereby confirming the robustness and scalability of the proposed design.

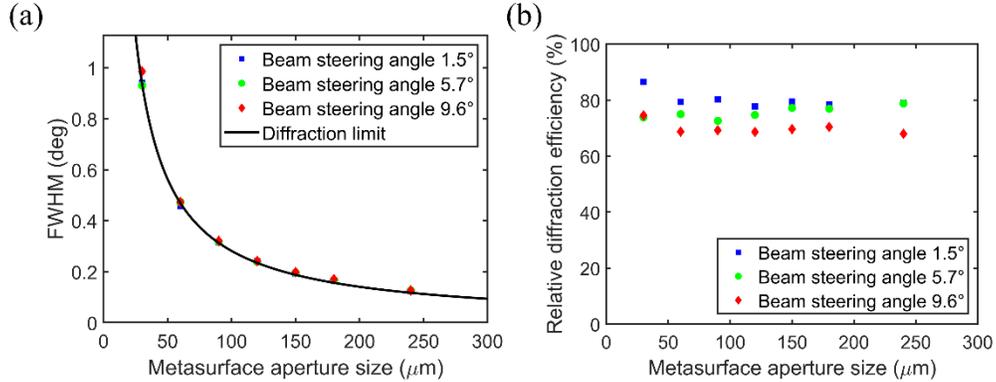

Fig. 3. (a) Angular FWHM of the primary diffraction lobe as a function of the aperture size, defined as the edge length of a square aperture. The black curve represents the diffraction-limited value. Here, we adopted the theoretical result under the small-angle approximation since the deviation introduced by the $1/\cos\theta_x$ factor remains below 1.5% for $\theta_x \leq 9.6°$. (b) Relative diffraction efficiency of the primary diffraction order as a function of the aperture size. In both panels, the blue squares, green circles, and red diamonds correspond to beam steering angles of $\theta_x = 1.5°, 5.7°$, and 9.6°, respectively.

## 3.3 Comparison with SLM-only design

An alternative approach to beam steering is to directly employ SLMs or other phase modulation platforms, such as silicon photonic phased arrays [33], to shape the phase distribution of the incident beam. In such systems, the phase profile can be optimized (apodized) to enhance beam quality [30]. Here, we compare our coherent-control-based design with this conventional SLM-only strategy and demonstrate the superior performance of our approach. Figure 4 plots the relative diffraction efficiency obtained from both methods. For a fair comparison, we assume the same SLM pixel pitch of 3 μm in both cases. In the SLM-only configuration, the phase and amplitude of each pixel are optimized using the same direct search algorithm applied in our design. The results reveal that our proposed design maintains a higher relative diffraction efficiency across a substantially broader angular range than the SLM-only configuration. This improvement originates from the tunable phase gradient within each supercell, which introduces an additional degree of control. In the absence of the metasurface, the maximum aliasing-free steering angle for the SLM with a 3 μm pixel pitch is restricted to $\arcsin(\lambda_0/2\Lambda)$ = 5.3°, where $\Lambda$ denotes the supercell period [34]. Consequently, the additional SLMs incorporated in the proposed configuration (Fig. 1b) do not replace the tunable metasurface with coherent control. Instead, the SLM and metasurface operate in a synergistic manner to enable large-aperture, high-quality beam steering across an extended angular range of approximately 11°.

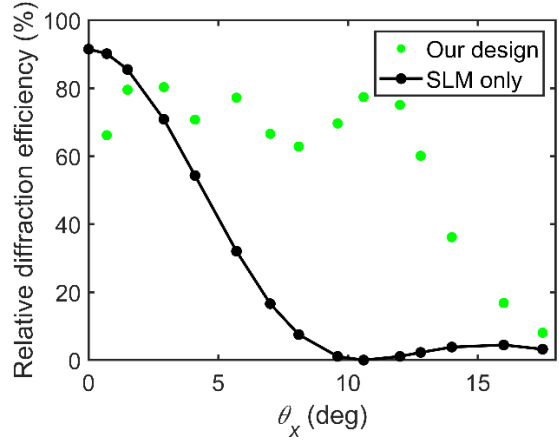

Fig. 4. Relative diffraction efficiency of the primary diffraction order: the proposed design vs. the SLM-only baseline. The green circles correspond to the results of the combined SLM and metasurface configuration, while the black circles and line indicate the SLM-only case.

### 3.4 2D beam steering

While the metasurface illustrated in Fig. 1a is designed with a phase gradient only along the $x$ direction, the direct search algorithm can be readily generalized to steer the beam across the full $(\theta_x, \theta_y)$ plane. This is achieved by allowing the tunable parameters of each supercell, namely the phase offset ($\varphi_0$) and the magnetic field ratio ($B_s/B_f$), to vary along both the $x$ and $y$ directions across the metasurface. Using the same FOM defined in Section 2.2, we calculated the relative diffraction efficiency in the case of 2D beam steering using direct search optimization, shown by the color scale plot in Fig. 5, as a function of $\theta_x$ and $\theta_y$.

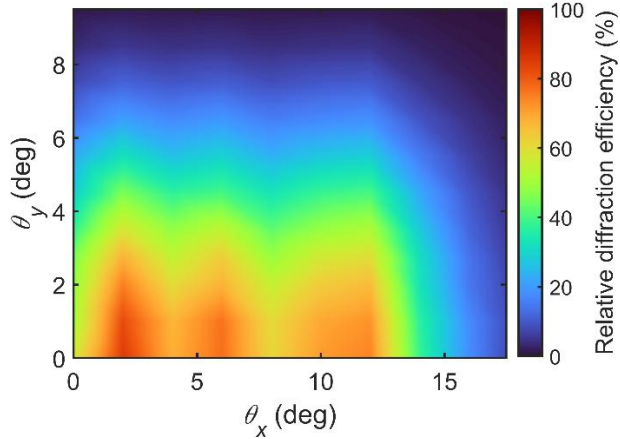

Fig. 5. Colormap showing the relative diffraction efficiency of the primary diffraction order as a function of $\theta_x$ and $\theta_y$, displayed with shading interpolation. The corresponding color scale is shown on the right. Results are presented for a square aperture with an edge length of 30 μm.

As observed in Fig. 5, the beam can be steered in both the $\theta_x$ and $\theta_y$ directions with high relative diffraction efficiency when the steering angles are small. However, the angular range of efficient steering in the $\theta_y$ direction is more limited than that in the $\theta_x$ direction by the aliasing-free condition, for the same reason as in the SLM-only configuration. This result is expected, since the metasurface does not provide a tunable linear phase gradient along the *y* direction. Despite this limitation, the proposed design remains promising for applications such as light detection and ranging (LiDAR), where a wider field of view is typically required only along one axis [35].

## 4. Conclusion

In this work, we have introduced a scalable approach that lifts the aperture-size constraint in coherent control of phase-gradient metasurfaces. The concept was validated through numerical modeling of continuous beam steering in large-area metasurfaces, demonstrating diffraction-limited beamforming over an ~11° field of view without aliasing. For a 150 μm × 150 μm metasurface comprising $2.5 \times 10^5$ meta-atoms, the scheme requires only SLMs with 2,500 independently controlled pixels, representing a substantial simplification compared with per-meta-atom tuning. Moreover, the proposed framework is inherently extensible to even larger-aperture metasurface platforms and other classes of tunable phase-gradient devices, underscoring its potential for a broad range of future photonic applications.

## Funding

Singapore-MIT Alliance for Research and Technology (SMART): Wafer-scale Integrated Sensing Devices based on Optoelectronic Metasurfaces (WISDOM) Interdisciplinary Research Group; UK Engineering and Physical Sciences Research Council (Grant No. EP/Z535813/1).

## Disclosures

The authors declare no conflicts of interest.

## Data availability statement

Data may be obtained from the authors upon reasonable request.

## References

1. P. P. Iyer, R. A. Decrescent, T. Lewi, N. Antonellis, and J. A. Schuller, "Uniform Thermo-Optic Tunability of Dielectric Metalenses," Phys Rev Appl **10**(4), (2018).
2. K. Zangeneh Kamali, L. Xu, N. Gagrani, H. H. Tan, C. Jagadish, A. Miroshnichenko, D. Neshev, and M. Rahmani, "Electrically programmable solid-state metasurfaces via flash localised heating," Light Sci Appl **12**, 40 (2023).
3. C. Damgaard-Carstensen, T. Yezekyan, M. L. Brongersma, and S. I. Bozhevolnyi, "Highly Efficient, Tunable, Electro-Optic, Reflective Metasurfaces Based on Quasi-Bound States in the Continuum," ACS Nano **19**(12), 11999–12006 (2025).
4. P. C. Wu, R. A. Pala, G. Kafaie Shirmanesh, W. H. Cheng, R. Sokhoyan, M. Grajower, M. Z. Alam, D. Lee, and H. A. Atwater, "Dynamic beam steering with all-dielectric electro-optic III–V multiple-quantum-well metasurfaces," Nat Commun **10**, 3654 (2019).
5. Y.-W. Huang, H. W. H. Lee, R. Sokhoyan, R. A. Pala, K. Thyagarajan, S. Han, D. P. Tsai, and H. A. Atwater, "Gate-Tunable Conducting Oxide Metasurfaces," Nano Lett **16**(9), 5319–5325 (2016).
6. B. Zeng, Z. Huang, A. Singh, Y. Yao, A. K. Azad, A. D. Mohite, A. J. Taylor, D. R. Smith, and H. T. Chen, "Hybrid graphene metasurfaces for high-speed mid-infrared light modulation and single-pixel imaging," Light Sci Appl **7**, 51 (2018).
7. O. Buchnev, N. Podoliak, M. Kaczmarek, N. I. Zheludev, and V. A. Fedotov, "Electrically Controlled Nanostructured Metasurface Loaded with Liquid Crystal: Toward Multifunctional Photonic Switch," Adv Opt Mater **3**(5), 674–679 (2015).

8. A. Komar, Z. Fang, J. Bohn, J. Sautter, M. Decker, A. Miroshnichenko, T. Pertsch, I. Brener, Y. S. Kivshar, I. Staude, and D. N. Neshev, "Electrically tunable all-dielectric optical metasurfaces based on liquid crystals," Appl Phys Lett **110**(7), 071109 (2017).
9. S. Q. Li, X. Xu, R. M. Veetil, V. Valuckas, R. Paniagua-Domínguez, and A. I. Kuznetsov, "Phase-only transmissive spatial light modulator based on tunable dielectric metasurface," Science (1979) **364**(6445), 1087–1090 (2019).
10. Z. Zhu, P. G. Evans, R. F. Haglund, and J. G. Valentine, "Dynamically Reconfigurable Metadevice Employing Nanostructured Phase-Change Materials," Nano Lett **17**(8), 4881–4885 (2017).
11. C. C. Popescu, K. Aryana, P. Garud, K. P. Dao, S. Vitale, V. Liberman, H. Bin Bae, T. W. Lee, M. Kang, K. A. Richardson, M. Julian, C. A. R. Ocampo, Y. Zhang, T. Gu, J. Hu, and H. J. Kim, "Electrically Reconfigurable Phase-Change Transmissive Metasurface," Advanced Materials **36**(36), 2400627 (2024).
12. Z. Fang, R. Chen, J. E. Fröch, Q. A. A. Tanguy, A. I. Khan, X. Wu, V. Tara, A. Manna, D. Sharp, C. Munley, F. Miller, Y. Zhao, S. Geiger, K. F. Böhringer, M. S. Reynolds, E. Pop, and A. Majumdar, "Nonvolatile Phase-Only Transmissive Spatial Light Modulator with Electrical Addressability of Individual Pixels," ACS Nano **18**(17), 11245–11256 (2024).
13. Y. Zhang, C. Fowler, J. Liang, B. Azhar, M. Y. Shalaginov, S. Deckoff-Jones, S. An, J. B. Chou, C. M. Roberts, V. Liberman, M. Kang, C. Ríos, K. A. Richardson, C. Rivero-Baleine, T. Gu, H. Zhang, and J. Hu, "Electrically reconfigurable non-volatile metasurface using low-loss optical phase-change material," Nat Nanotechnol **16**, 661–666 (2021).
14. E. Arbabi, A. Arbabi, S. M. Kamali, Y. Horie, M. Faraji-Dana, and A. Faraon, "MEMS-tunable dielectric metasurface lens," Nat Commun **9**(1), 812 (2018).
15. C. Zhang, J. Jing, Y. Wu, Y. Fan, W. Yang, S. Wang, Q. Song, and S. Xiao, "Stretchable All-Dielectric Metasurfaces with Polarization-Insensitive and Full-Spectrum Response," ACS Nano **14**(2), 1418–1426 (2020).
16. X. Fang, K. F. MacDonald, and N. I. Zheludev, "Controlling light with light using coherent metadevices: all-optical transistor, summator and invertor," Light Sci Appl **4**(5), e292–e292 (2015).
17. J. Shi, X. Fang, E. T. F. Rogers, E. Plum, K. F. MacDonald, and N. I. Zheludev, "Coherent control of Snell's law at metasurfaces," Opt Express **22**(17), 21051 (2014).
18. X. Fang, M. Lun Tseng, J.-Y. Ou, K. F. MacDonald, D. Ping Tsai, and N. I. Zheludev, "Ultrafast all-optical switching via coherent modulation of metamaterial absorption," Appl Phys Lett **104**(14), 141102 (2014).
19. M. Papaioannou, E. Plum, J. Valente, E. T. Rogers, and N. I. Zheludev, "Two-dimensional control of light with light on metasurfaces," Light Sci Appl **5**(4), e16070–e16070 (2016).
20. X. Fang, M. L. Tseng, D. P. Tsai, and N. I. Zheludev, "Coherent Excitation-Selective Spectroscopy of Multipole Resonances," Phys Rev Appl **5**, 014010 (2016).
21. S. Yin, F. He, W. Kubo, Q. Wang, J. Frame, N. G. Green, and X. Fang, "Coherently tunable metalens tweezers for optofluidic particle routing," Opt Express **28**(26), 38949–38959 (2020).
22. F. He, K. F. MacDonald, and X. Fang, "Continuous beam steering by coherent light-by-light control of dielectric metasurface phase gradient," Opt Express **28**(20), 30107–30116 (2020).
23. F. He, Y. Feng, H. Pi, J. Yan, K. F. MacDonald, and X. Fang, "Coherently switching the focusing characteristics of all-dielectric metalenses," Opt Express **30**(15), 27683–27693 (2022).
24. F. Yang, K. P. Dao, S. An, X. Qiu, Y. Zhang, J. Hu, and T. Gu, "Design of continuously tunable varifocal metalenses," Journal of Optics **25**(11), 115102 (2023).
25. M. Y. Shalaginov, S. D. Campbell, S. An, Y. Zhang, C. Ríos, E. B. Whiting, Y. Wu, L. Kang, B. Zheng, C. Fowler, Hualiang. Zhang, D. H. Werner, J. Hu, and T. Gu, "Design for quality: reconfigurable flat optics based on active metasurfaces," Nanophotonics **9**(11), 3505–3534 (2020).
26. S. Shrestha, A. C. Overvig, M. Lu, A. Stein, and N. Yu, "Broadband achromatic dielectric metalenses," Light Sci Appl **7**(1), 85 (2018).
27. W. T. Chen, A. Y. Zhu, V. Sanjeev, M. Khorasaninejad, Z. Shi, E. Lee, and F. Capasso, "A broadband achromatic metalens for focusing and imaging in the visible," Nat Nanotechnol **13**(3), 220–226 (2018).
28. F. Presutti and F. Monticone, "Focusing on Bandwidth: Achromatic Metalens Limits," Optica **7**(6), 624–631 (2020).
29. D. Chen, J. Yang, J. Huang, Z. Zhang, W. Xie, X. Jiang, X. He, Y. Han, Z. Zhang, and Y. Yu, "Continuously tunable metasurfaces controlled by single electrode uniform bias-voltage based on nonuniform periodic rectangular graphene arrays," Opt Express **28**(20), 29306–29317 (2020).
30. P. Thureja, G. K. Shirmanesh, K. T. Fountaine, R. Sokhoyan, M. Grajower, and H. A. Atwater, "Array-level inverse design of beam steering active metasurfaces," ACS Nano **14**(11), 15042–15055 (2020).

31. F. Yang, S. An, M. Y. Shalaginov, H. Zhang, C. Rivero-Baleine, J. Hu, and T. Gu, "Design of broadband and wide-field-of-view metalenses," Opt Lett **46**(22), 5735 (2021).
32. W. T. Chen, J. S. Park, J. Marchioni, S. Millay, K. M. A. Yousef, and F. Capasso, "Dispersion-engineered metasurfaces reaching broadband 90% relative diffraction efficiency," Nat Commun **14**, 2544 (2023).
33. X. Sun, L. Zhang, Q. Zhang, and W. Zhang, "Si Photonics for Practical LiDAR Solutions," Applied Sciences **9**(20), 4225 (2019).
34. T. Gu, H. J. Kim, C. Rivero-Baleine, and J. Hu, "Reconfigurable metasurfaces towards commercial success," Nat Photonics **17**(1), 48–58 (2023).
35. Z. Dai, A. Wolf, P. P. Ley, T. Glück, M. C. Sundermeier, and R. Lachmayer, "Requirements for Automotive LiDAR Systems," Sensors **22**(19), 7532 (2022).